\documentclass[11pt]{article}
\usepackage{moriond,epsfig}
\usepackage{amsmath,amssymb}

\def\be{\begin{equation}}
\def\ee{\end{equation}}
\def\bea{\begin{eqnarray}}
\def\eea{\end{eqnarray}}

\begin{document}
\vspace*{-0.5cm}
\title{\Large Towards Inflation and Accelerating Cosmologies\\
in String-Generated Gravity Models}

\author{\bf Ishwaree P. Neupane}

\address{Department of Physics and Astronomy, University of Canterbury,\\
Private Bag 4800, Christchurch 8004, New Zealand \\ and\\
Central Department of Physics, Tribhuvan University\\
Kirtipur, Kathmandu, Nepal}

\maketitle\abstracts{The string $\alpha^\prime$-correction to the
usual Einstein action comprises a Gauss-Bonnet integrand
multiplied by non-trivial functions of the modulus field $\chi$
and/or the dilaton field $\phi$. We discuss how the presence of
such terms in the four dimensional effective action can explain
several novel phenomena, such as a four-dimensional flat
Friedmann-Robertson-Walker universe undergoing a cosmic inflation
at the early epoch, as well as a cosmic acceleration at late
times. The model predicts, during inflation, spectra of both
density perturbations and gravitational waves that may fall well
within the experimental bounds. The model therefore provides a
unified approach for explaining the early and late time
accelerating phases of the universe. }


String/M theory has the prospect of unification of all
interactions (see~\cite{Dienes:1996du} for a review), providing
inter-relationships between quantum mechanics and general
relativity. The hope is that the ultimate theory of Nature
contains a fundamental and simple, concise relation between
close-by and far-away, and past and future. However, string
theory, in which great progress was made in the 80's and 90's, has
not been developed to the stage where the full theory may be
employed to construct a detailed cosmology, explaining the
evolution of our universe from an epoch near the Planck scale
($\sim {10}^{-33}\,{\rm cm}$) to the presently accessible largest
scale ($\sim {10}^{28}\,{\rm cm}$).

The assumptions which make string cosmology a challenging subject
are the existence of (yet unobserved) extra dimensions and
supersymmetry. Modern (super)string theory or M theory has
additional features: it now goes beyond the perturbative
approach~\cite{Candelas:1985,Gross:1984dd} popularised in the
$80$'s and $90$'s and takes into account the effects due to
fluxes, branes, singularities and non-trivial cycles in the extra
dimensions. M theory provides new avenues for studying the theory
of inflation or cosmic acceleration via flux~\cite{Kachru:2003} or
time-dependent~\cite{Ish03} compactifications.

The heterotic string theory after its discovery in
1985~\cite{Candelas:1985,Gross:1984dd} appeared to be the string
theorist's dream as something like the real world. Only in 1995,
there came a breakthrough, not in finding the right version to
describe the physical world but in understanding the connections
among various versions of string theory, which are related to each
other by various string dualities~\cite{Sen:1996a}. It is
therefore not necessary that all five versions of string theory to
be equally relevant in a cosmological context. The
four-dimensional effective action of our interest here is given by
\begin{equation}\label{dilatonGB2} S = \int
d^{4}{x}\sqrt{-g}\bigg[ \frac{1}{2\kappa^2} R -{\gamma}\, (\Delta
\chi)^2-V(\chi)-\zeta\, (\Delta\phi)^2 +
\frac{1}{8}[\lambda(\phi)- \xi(\chi)] {\cal R}^2_{GB} \bigg],
\end{equation}
where $\kappa$ is the inverse Planck mass $m_{\rm Pl}^{-1}=(8\pi
G_N)^{1/2}$ and ${{\cal R}^2_{GB}} \equiv  R^2-4 R_{\mu\nu}
R^{\mu\nu} + R_{\mu\nu\rho\chi} R^{\mu\nu\rho\chi}$ is the
Gauss-Bonnet (GB) integrand. Here $\phi$ may be viewed as a
dilaton, which has no potential, whereas $\chi$ is an ordinary
scalar (or inflaton field) which, in particular cases, could be a
modulus field associated with the overall size of extra
dimensions. In a known example of heterotic string
compactification~\cite{Narain92a,Antoniadis93a},
$\lambda(\phi)=\lambda_1\,e^\phi+\cdots$ and $\xi(\chi)\propto
(\ln (2) -\frac{2\pi}{3}\cosh(\chi))+\cdots$ ($\chi$ is the real
part of the common volume modulus). Unfortunately, we do not have
a precise knowledge about $V(\chi)$; any such potential may take
into account the supersymmetry breaking non-perturbative
effects~\cite{Kachru:2003}. It may well be that the potential is a
function of both $\chi$ and $\phi$, so $V(\chi)\to V(\chi,\phi)$.

Nojiri et al.~\cite{Nojiri:2005vv} examined a special (exact)
solution for the system~(\ref{dilatonGB2}) with the choice
$V(\chi)=V_0\, {\rm e}^{-\chi/\chi_0}$ and $\xi(\chi)=\xi_0 {\rm
e}^{\chi/\chi_0}$,
deleting the $\phi$-dependent terms. The analysis was extended
in~\cite{Sami:2005zc} (see also~\cite{Calcagni:2005}) by
introducing higher (than second order) curvature corrections, but
in the $\chi$=const background. A more general solution, in a
fixed dilaton background, $\dot{\phi}=0$, was given
in~\cite{Ish:05d}. It was explained there how the model
(\ref{dilatonGB2}) can explain a relaxation of vacuum energy (or
effective potential) to a small value (exponentially close to
zero) after a sufficiently large number of e-folds of expansion,
providing a possible solution to both inflation and cosmological
constant problems.

For simplicity, we shall start our discussion by neglecting all
the dilaton related terms. We find it convenient to define the
following variables (in the units $\kappa=1$):
\begin{equation}\label{new-variables}
X\equiv {\gamma}\, ({\dot{\chi}}/{H})^2, \quad Y\equiv
{V(\chi)}/{H^2}, \quad U\equiv \xi(\chi) H^2 \quad \varepsilon
\equiv {\dot{H}}/{H^2}={H^\prime}/{H},
\end{equation}
where the dot (prime) denotes a derivative w.r.t. the proper time
$t$ (number of e-folds, $N=\int H\, {d t}=\ln (a(t)/a_0)$). Note
that $N$ is a monotonically increasing function of the scale
factor, $a(t)$, or the proper time $t$, but its sign depends on
the assumption of what the scale $a_0$ represents. If $a_0$ is the
present value of the scale factor, then $N$ is negative in the
past, $a(t)<a_0$.

In the absence of the Gauss-Bonnet coupling (so $U=0$), the
equations of motion are given by $Y =3+\varepsilon$ and
$X=-\varepsilon$. These are remarkably simple! However, different
choice of $\varepsilon$ implies different $Y$ and hence different
$V(\chi)$, and so is the equation of state (EOS) parameter $w$,
which is defined by $ w=-2\varepsilon/{3}-1$. Is it then all
worthy?

In the past, many authors have considered cosmological solutions
with various {\it ad hoc} choices of the potential, e.g. chaotic
potential $V(\chi)= V_0 + \frac{1}{2} m^2\chi^2$, exponential
potential $V(\chi)= V_0\,e^{- p\, \chi}$, inverse power-law
potential $V(\chi)=\Lambda^4 (\chi_0/\chi)^n$ ($n\geq 2$), the
axion-potential $V(\chi)= \Lambda^4 (c \pm \cos(\chi/\chi_0))$,
and so on. The models with exponential and inverse power-law
potentials possess some interesting features at late times, such
as the cosmological attractors~\cite{Ferreira97a,Steinhardt:98a}.
Though it may be desirable to construct a cosmological model that
gives rise to
\begin{equation} w\approx -1,
\end{equation}
for the model to work, the field $\chi$ must relax its potential
energy after inflation down to a sufficiently low value, possibly
very close to the present value of {\it dark energy},
$\Lambda_{vac}\sim (10^{-3}\,{\text{eV}})^4$. Do all of the above
potentials possess such feature? The answer is soundly no! An
exponential potential is perhaps the best choice but it is more
plausible that the coupling constant $p$ is a function of the
field $\chi$ itself, or the number of e-folds, $N\equiv
|\ln(a_f/a_i)|$.

For generality, we assume that $\xi(\chi)\neq 0$, without
specifying its sign at this stage. In an accelerating universe,
the Gauss-Bonnet term is positive, i.e., ${\cal
R}_{GB}^2=24\frac{\dot{a}^2\ddot{a}}{a^3}>0$. The effective
potential, which may be given by $\Lambda(\chi)\equiv
V(\chi)-\frac{1}{8}\, \xi(\chi) R_{GB}^2$, can be (exponentially)
small for $V(\chi)\approx \frac{1}{8}\, \xi(\chi) R_{GB}^2$. We
find it convenient to impose the condition~\cite{Ish:05d,Ish:05e}
\begin{equation}\label{gauge-choice}
\Lambda(\chi)\equiv H^2 (3+\varepsilon),
\end{equation}
so that $\Lambda(\chi)\ge 0$ for $\varepsilon\ge -3$ or
equivalently when the EOS parameter $w \le 1$. There is now only
one free parameter in the model, which may be fixed either by
allowing one of the variables in (\ref{new-variables}) to take a
fixed (but arbitrary) value or by making an appropriate ansatz.
Here we consider two physically motivated cases. First consider
the case of power-law inflation, for which
\begin{equation}\label{scale-factor}
a(t) \equiv e^{\omega(t)} \equiv \left(c_0 t+
t_1\right)^{-1/\varepsilon_0},
\end{equation}
where $c_0$ and $t_1$ are arbitrary constants. This actually
implies that $\varepsilon\equiv \dot{H}/H^2=\varepsilon_0$. For
$\varepsilon_0 < 0$, $c_0>0$ and $t_1$ may be set zero, but, for
$\varepsilon_0>0$, one may take $c_0<0$ and set $t_1=t_\infty$,
just to ensure that the scale factor grows with $t$ in either
case. With (\ref{scale-factor}), we find $H\propto
e^{\varepsilon_0 N}$ and $\xi(\chi)H^2 =c_2\,e^{\alpha_+
N}+c_3\,e^{\alpha_- N}$, where $\alpha_\pm =
\frac{1}{2}\left(\varepsilon_0-5 \pm \sqrt{
9(\varepsilon_0+3)^2-32}\right)$. For the reality of $\alpha_\pm$,
it is required that $\varepsilon_0>-1.11$ or
$\varepsilon_0<-4.88$; we rule out the latter as it implies $w>1$.

\begin{figure}[ht]
\begin{center}
\epsfig{figure=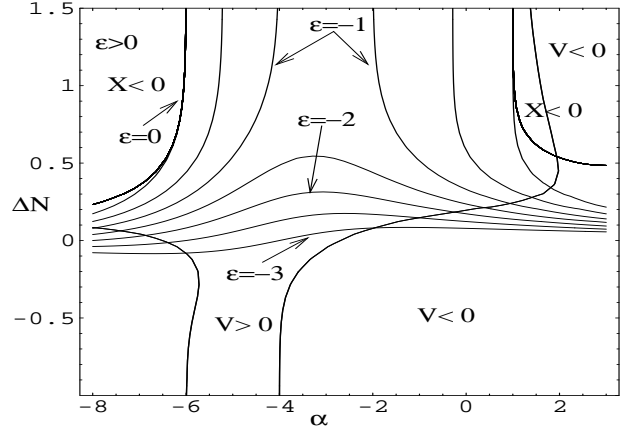,height=2.3in,width=3.2in}
\end{center}
\caption{The $\Delta N$ - $\alpha$ phase space. The almost
horizontal line with $\varepsilon=-3$ separates the regions
between $\Lambda(\chi)>0$ and $\Lambda(\chi)< 0$. Here we have set
${\alpha_0}=\ln(0.5)$. The region with $\varepsilon<-3$ (or
$\Lambda(\chi) <0$) is physically less relevant as it implies
$w>1$. Only the lines with $\varepsilon>-1$ fall in the
accelerating region. For $-6\le \alpha\le -4$, $V(\chi)$ is always
non-negative; $V(\sigma)\ge 0$ may be required in order that the
vacuum becomes a possible ground state. } \label{figure1}
\end{figure}

Spurred on by this example, we make no assumption about the form
of scale factor, and instead consider the ansatz that during a
given epoch we may make the approximation
\begin{equation}\label{ansatz1}
U \equiv \xi(\chi) H^2 \equiv e^{\alpha_0+ \alpha N},
\end{equation}
where $\alpha_0$ and $\alpha$ are arbitrary at this stage. This
allows us to write $\varepsilon$ in a closed form:
\begin{equation}\label{main-sol}
\varepsilon= \beta \tanh\beta (N+N_1)-\hat{\beta} \quad
\Rightarrow H\propto \cosh\beta (N+N_1)\,e^{- \hat{\beta}\, N},
\end{equation}
where $N_1$ is a free parameter, $\hat{\beta} \equiv
(16+\alpha)/{4}$ and $\beta \equiv \frac{1}{4}\,\sqrt{
9\alpha(\alpha+8)+208}$. Initially, $N$ may take a large negative
value, $N\equiv \ln (a/a_0)\ll 0$. As we wish to obtain an
inflationary epoch with sufficiently large number of e-folds of
expansion, say $65$ e-folds, it is reasonable to take $N_1\sim
{\cal O}(65)$. For $\Delta N\equiv N+N_1 \lesssim 0$, the universe
is not accelerating, but it starts to accelerate
($\varepsilon>-1$) for $\Delta N\gtrsim 0.2$. The solution where
$-6<\alpha<-4$ (or $-2<\alpha<1$) is accelerating with the EOS
parameter $w\ge -1$; of course, the solution where $\alpha<-6$ (or
$\alpha>1$) is also accelerating but in this case, since
$\varepsilon>0$ or $X<0$ (cf figure~\ref{figure1}), $\chi$ may
behaves as a phantom field.

The scalar field potential may be given by
\begin{equation} \label{main-sca-po}
V(\chi)=H^2\left(3+\varepsilon+3(\varepsilon+1)
U\right).
\end{equation}
Note that at the start of inflation, $\varepsilon\approx -1$ (or
$w\approx -1/3$), the GB term is not contributing; inflation can
be mainly due to the potential. One may express
(\ref{main-sca-po}) in terms of the scalar field $\chi$, although
it involves a somewhat complicated relation between $N$ and
$\chi$, namely,
\begin{eqnarray}
& \gamma{\chi^\prime}^{\,2}=
-\frac{1}{2}\left[\alpha(\alpha-1)-2({\hat\beta}^2+\beta^2)+(\alpha-2)\hat{\beta}
+(2\beta-\alpha\beta+4\hat{\beta}\beta)\tanh(\beta\Delta N)\right] U\nonumber \\
& \quad +\, \hat{\beta}-\beta\tanh(\beta\Delta N).
\end{eqnarray}
After a certain number of e-folds of expansion, i.e., $\beta\Delta
N \gtrsim 2$, this simplifies to
\begin{equation}
\gamma{\chi^\prime}^{\,2}\to
\frac{e^{\alpha_0}}{2}(\alpha+2\hat{\beta}-2\beta)(1-\alpha+\hat{\beta}-\beta)
e^{\alpha N}+\hat{\beta}-\beta \equiv x_0(\alpha)
\end{equation}
Especially, for $\alpha=1$, we get $\hat{\beta}=\beta$ and hence
$\chi^\prime\simeq 0$ when $\beta\Delta N\gtrsim 2$, i.e., the
time-variation of the field $\chi$ or its rolling with $N$ is
negligibly small. In this case $\Lambda(\chi)$ acts purely as a
cosmological constant term. Note that $\hat{\beta}=\beta$ is
obtained also for $\alpha=-6$, but in this case $\chi$ may behave
as a phantom field unless $\xi(\chi)$ is chosen to be negative or
the coefficient $e^{\alpha_0}$ is exponentially small, $\alpha_0
\to -\infty$. In the limit $U\ll 1$, so that the Gauss-Bonnet
correction is small and is not dominating the dynamics, the
potential is well approximated by $V(\chi)\propto H^2 \propto
e^{2(\beta-\hat{\beta})N}\equiv V_0 \, e^{-p\, \chi}$ where
$p\equiv 2(\hat{\beta}-\beta)\lambda_0$ and
$\lambda_0=\sqrt{\frac{\gamma}{x_0(\alpha)}}=
|\chi^\prime|_{\beta\Delta N>2}$. Furthermore, we get
$\Lambda(\chi)=\Lambda_0\,e^{-p\,\chi}$, which is valid for any
value of $\alpha$ (or $N$) but satisfying $\beta\Delta N
>2$.

For a canonically normalized field $\chi$ such that
$K(\chi)=\frac{1}{2}\dot{\chi}^{\,2}$, the exponential potential
$\Lambda=\Lambda_0\,e^{-p\,\chi}$ with $p=p(\chi)$ possesses some
interesting cosmological features~\cite{Tashiro:03qp}, which are
unavailable if $p$ is a constant. Notably, one would require
$p<\sqrt{2}$ to obtain an inflationary solution, where as
$p>\sqrt{6}$ may be required after inflation in order for the
potential energy not to dominate the energy content of the
universe, satisfying the necleosynthesis bound, namely $p^2
> n/{\Omega_\chi^{\rm max}}$, where $n=3$ ($4$) for matter
(radiation). The nucleosysthesis bound is not violated for
$(\hat{\beta}-\beta)\lambda_0 \gtrsim \sqrt{5}$, given that
$\Omega_\chi^{\rm max}\lesssim 0.2$. Though the model studied
in~\cite{Barreiro:99}, namely, $V(\chi)=\Lambda^4 \left(e^{-\,P
\chi}+e^{-\,Q \chi}\right)$, did not work satisfactorily, as it
required $P>5.5$ (and $Q<0.8$) in order to satisfy a constraint
coming from the nucleosynthesis bounds on $\Omega_\chi$, a scalar
potential as a sum of two (or more) exponential terms of different
slopes may be interesting from some other reasons, for example, it
gives rise to more than one period of cosmic
acceleration~\cite{Barreiro:99,Ish:03-05}.

\begin{figure}[ht]
\begin{center}
\epsfig{figure=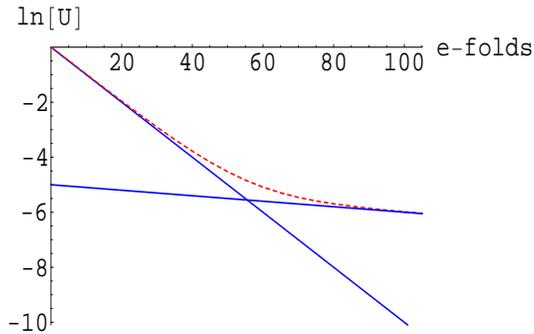,height=2.1in,width=2.8in}
\end{center}
\caption{The qualitative behavior of the product
$\xi(\chi)H^2\equiv U(\chi)= e^{\alpha_0+\alpha N} +
e^{\alpha_1+\hat{\alpha} N}$ (dotted line) as a function of the
number of e-folds, with $\alpha=-0.1$, $\hat{\alpha}=-0.01$,
$\alpha_0=0$ and $\alpha_1=-5$. The time-variation of the coupling
function $U(\chi)$ may be large at early times (before the end of
inflation), but it is nearly constant at late times.}
\label{figure2}
\end{figure}

Let us now return to the discussion related to (\ref{main-sol}).
The scale factor of the universe after inflation would naturally
become much larger than its initial value, but it is smaller than
$a_0$, i.e. the present value, so $-{\cal O}(1) \lesssim N<0$. Our
approximation (\ref{ansatz1}) that $U$ be given by a single
exponential term may break down at some intermediate epoch (cf
Fig~\ref{figure2}). Sometime later, $U$ may be approximated by the
term $\propto e^{\hat{\alpha} N}$. The solution is then given by
(\ref{main-sol}), with $\alpha$ and $N_1$ replaced by
$\hat{\alpha}$ and $N_0$, respectively. For $N+N_0 \lesssim 0$,
the universe is in a deceleration phase which implies that
inflation must have stopped during the intermediate epoch. As $N$
crosses $N_{0}$, or when $N+N_0>0$, the universe may begin to
accelerate for the second time.

For $a(t)\sim a_0$, we get $N\sim 0$ and hence $\Delta \chi
\propto \Delta N$; the constant of proportionality may be small.
This implies that, after inflation, even if $\hat{\alpha}$ is very
different from unity, the time-variation of $\chi$ may be small,
unless we do not demand a large number of e-folds of expansion. It
is also possible that after inflation the field $\chi$ stabilises
(i.e. no significant rolling with $N$) and the late time
acceleration is due to slow rolling of the field $\phi$. Later we
will discuss about this possibility. Below we discuss the result
of cosmological perturbations for a single scalar field $\chi$.

\begin{figure}[ht]
\hskip0.1in
\epsfig{figure=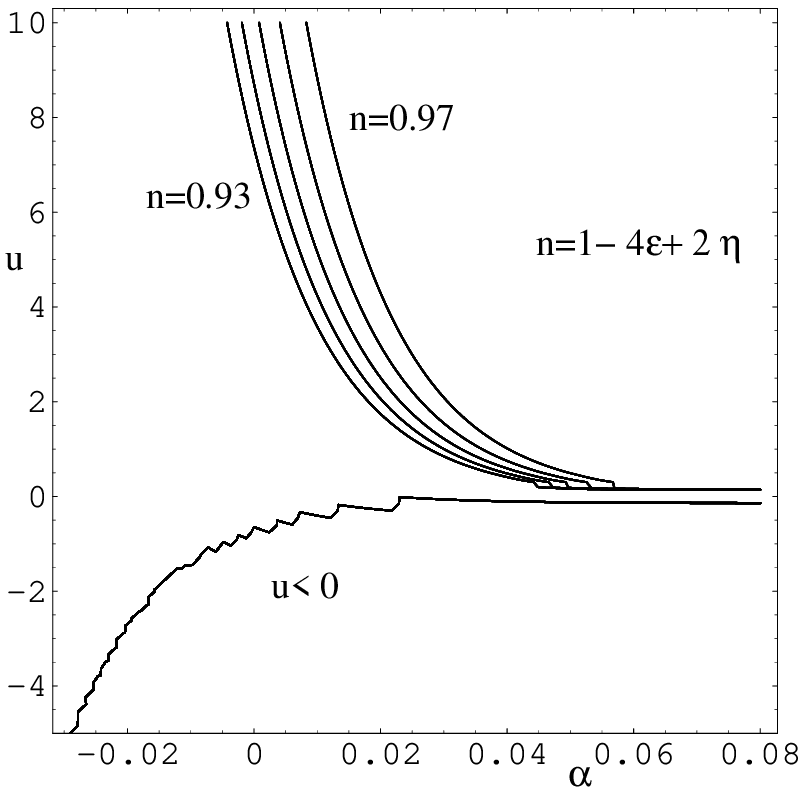,height=2.1in,width=2.7in}
\hskip0.5in
\epsfig{figure=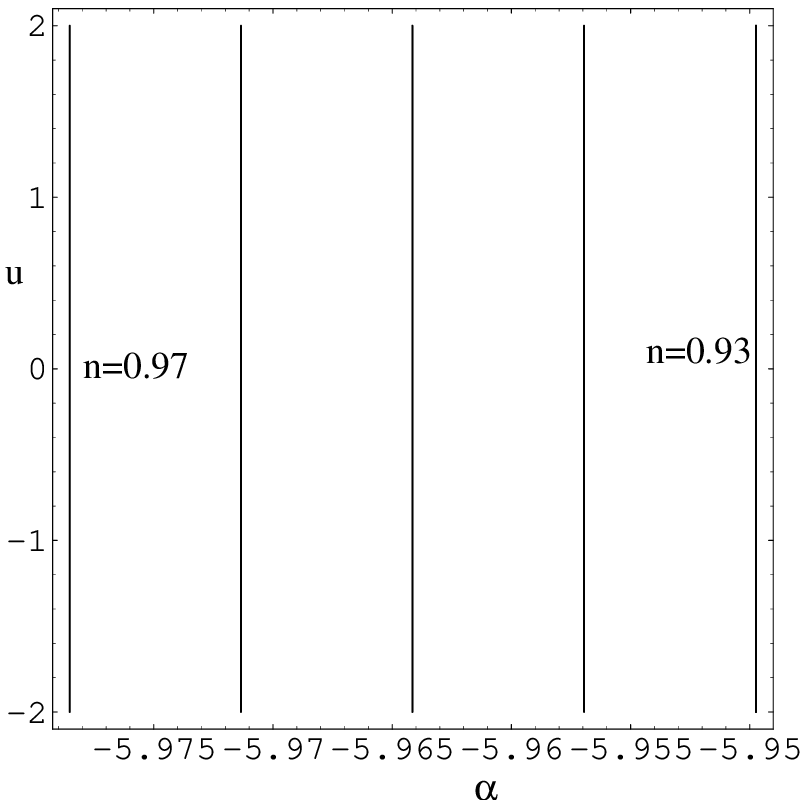,height=2.1in,width=2.7in}
\caption{The contour plot of $\alpha$ and $u$ ($\equiv
\exp(\alpha_0)$), showing the value of the spectral index
$n=1-4\epsilon+2\eta$ in the range $[0.93, 0.97]$; $n$ is
independent of the value of $u$ for a large negative value of
$\alpha$ ($<-4$) } \label{figure3}
\end{figure}

It is generally believed that during inflation the inflaton and
graviton field undergo quantum-mechanical
fluctuations~\cite{Linde-Others}, leading to scalar (density) and
tensor (gravity waves) fluctuations, which in turn would give rise
to significant effects on the large-scale structures of the
universe at the present epoch. In turn, it may be hoped that the
spectra of perturbations provide a potentially powerful test of
the inflationary hypothesis. In this respect, one may ask whether
the solution of the type (\ref{main-sol}) can generate scalar
perturbations of the desired magnitudes, like $\delta_H\equiv
\delta\rho/\rho \sim 2\times {10}^{-5}$ and $n \simeq 0.95$. The
spectral index (or tilt parameter) is given by
\begin{equation}
n \simeq 1-4 \epsilon + 2 \eta,
\end{equation}
where $\epsilon \equiv 2(\ln H)^\prime/{\chi^\prime}^{\,2} $ and
$\eta \equiv 2\left[H^{\prime\prime}/H - (\ln H)^\prime
(\ln\chi^{\prime})^{\prime}\right]/{\chi^\prime}^{\,2}$. We note
that inflationary predictions with $\xi(\chi)\neq 0$ can be
significantly different from those with $\xi(\chi)=0$
(see~\cite{Hwang97gf} for discussions on the related theme). In
our model, a spectral index of magnitude $n \simeq 0.95$ may be
obtained with $\alpha\simeq -5.964$ or with $\alpha \sim [-0.04,
0.04]$ (cf figure~\ref{figure3}). Furthermore, for $\alpha\approx
-6$ and $\alpha \gtrsim 0$, the tensor-to-scalar ratio may be
negligibly small, $r\equiv A_T/A_S \ll 0.4$~\cite{Ish:05e}.

The acceleration of the universe at late times~\cite{Bennett03a}
is a surprising and enormously influential result. It is
worthwhile to know if the dilaton plays a special role in the
present accelerated phase of the universe. So far we discussed the
result with no $\phi$-dependent terms in the action, or that
$\dot{\phi}\simeq 0$. In the presence of $\phi$-dependent terms,
one may introduce two more variables:
\begin{equation}
\Phi\equiv \lambda(\phi) H^2, \quad W\equiv \zeta\,
({\dot{\phi}}/{H})^2.
\end{equation}
First consider the case $\zeta=0$, for which there is no kinetic
term for $\phi$. $\Phi$ is assumed to be positive; of course, its
time variation must be small at late times so as not to conflict
with ground based solar system tests of GR. Here we do not impose
any condition like (\ref{gauge-choice}). Especially, for $U=\Phi$,
for which $\phi$ is behaving similarly as $\chi$ and hence
$f\equiv \lambda(\phi)-\xi(\chi)=0$, we find~\cite{Ish:05f}
\begin{equation}
X=-(\ln U)^\prime, \quad Y=3+ (\ln U)^\prime/2, \quad
\varepsilon=(\ln U)^\prime/2.
\end{equation} The solution $U\propto e^{\alpha N}$
leads to an accelerated expansion of the universe for $\alpha>-2$;
in particular, for $\alpha=-0.09$, we get $\varepsilon=-0.045$ and
hence $w=-0.97$. According to the recent WMAP result~\cite{WMAP3},
putting together all the experiments, the measured value of the
dark energy EOS parameter is $w_{DE}=-\,1.06_{-\,0.08}^{+\,0.13}$.
For the above solution, it is required that
$\alpha=0.18_{+\,0.24}^{-\,0.39}$. Especially, for $\alpha=0$, the
effect is similar to that of a cosmological constant term,
$w_{\Lambda}=-1$.

In the case $\zeta\neq 0$, the field equations are solved
explicitly, for example, with $U=\Phi={\rm const}\equiv \Phi_0$.
Then, there exist three branches of solution: the first branch is
given by $\varepsilon=\sqrt{6} \tanh (\sqrt{6}(N+N_0))-3$ and
$W=-3\Phi_0(1+2\varepsilon)$. For this solution the universe
enters into an accelerating phase for $N+N_0\gtrsim 0.22$. In
particular, for $N+N_0\gtrsim 2$, the deceleration parameter $q$
is almost constant, $q\equiv
-\,a\ddot{a}/{\dot{a}}^2=-1-\varepsilon\simeq 0-0.45$.
Furthermore, we get $W\simeq 0.3 \Phi_0$, so that, for $\Phi_0>0$,
$\phi$ behaves as a canonical scalar. Either both of the other two
branches support an accelerating phase with $w>-1$, or only one of
the branches supports inflation with $w<-1$.

Finally consider a situation near the analytic minimum where
$V(\chi)\simeq 0$. In this case a cosmic acceleration at late
times may arise due to slow rolling of the field $\phi$ and/or
$\xi$:
\begin{equation}
\dot{\chi}\equiv \sqrt{\frac{|\chi_0|}{\gamma}}\, H, \quad
\dot{\phi}=\sqrt{\frac{|\phi_0|}{\zeta}}\, H
\left(\frac{a_0}{a}\right)^{m/2}\equiv
\sqrt{\frac{|\phi_o|}{\zeta}}\, H\,e^{-m N/2},
\end{equation} where $\chi_0$, $\phi_0$ and $m$ are arbitrary at
this stage. The solution is given by
\begin{eqnarray}\label{two-scalars-zeroV}
&&\Xi \equiv \dot{\chi} H \frac{d\xi(\chi)}{d\chi}=
\frac{4\chi_0(\chi_0-3)+2\chi_0\phi_0(m-2)
e^{-m N}}{3\left(4\chi_0+(4-m)\phi_0\, e^{-m N} \right)},\nonumber \\
&& \Psi \equiv \dot{\phi} H \frac{d\lambda(\phi)}{d\phi}
=\frac{1}{3}\,\frac{(m+4)\chi_0\phi_0+(m-4){\phi_0}^2
\left(3-e^{-m N}\right)}
{4\chi_0\, e^{m N}+(4-m)\phi_0},\nonumber \\
&&\varepsilon=-1+\frac{(m-4)\phi_0 \,e^{-m N}-4 \chi_0}{3+\chi_0+
\phi_0\,e^{-m N}}.
\end{eqnarray}
The solution may be modelled such that $\ln(1+z)=|\Delta
N|=\ln(a_f/a_i)$, where $z$ is the red-shift factor. The number of
e-folds $\Delta N\sim 0.7$ when $z$ drops from $1$ to $0$. For
simplicity, let us assume that $\chi_0\simeq 0$, so that $\chi$
stabilises. Then the solution (\ref{two-scalars-zeroV}) can
support a transition between phantom-like phase ($\varepsilon>0$)
and standard (non-phantom) phase~\footnote{Such a transition is
reported to occur in variants of string-inspired gravitational
theories, see, e.g.,~\cite{McInnes05vp,Nojiri:2005jg,Aref'eva06},
and also in the model of quintessence (non-minimally) coupled to
an ordinary scalar or
dilaton~\cite{Perivola:2005,Wei:05si}.}($\varepsilon<0$),
especially, for $m\ge 5$. However, if $\phi$ stabilises
($\dot{\phi}=0$) before $\chi$ stabilises, then we find
$\varepsilon\simeq -1-4\chi_0/(3+\chi_0)$; with $\chi_0>0$, as
such the case for a canonical $\chi$, the solution is not
accelerating since $\varepsilon>-1$. However, if both scalars
$\phi$ and $\chi$ are (slowly) rolling with time, it is possible
that the EOS parameter $w<-1/3$, although its value is more
sensitive to the choice of the parameters $\chi_0$ and $\phi_0$.

In summary, we have shown that the string
$\alpha^\prime$-correction to the usual gravitational action of
general relativity of the Gauss-Bonnet form plays an important and
interesting role in explaining the early and late time
accelerating phases of the universe, with singularity-free
solutions. For constant dilaton phase of the standard scenario,
the model can predict during inflation spectra of both density
perturbations and gravitational waves that may fall well within
the experimental bounds. It is hoped that the results in this
paper help for explaining both the inflation and the cosmic
acceleration at late times. For completeness, in particular, in
order to study the transition between deceleration and
acceleration, one may have to include matter field which is also
the constituent that we know dominates the universe during
deceleration (see~\cite{Nojiri:06je} for a work in this
direction). In the presence of a barotropic fluid
(matter/radiation), one defines an effective EOS parameter,
$w_{eff}$, which may differ from the $w$ obeyed by a source of
{dark energy} alone, like the scalar potential, $V(\chi)$, or by
some exotic source of {dark energy} such as K-{\it inflaton}.

\medskip
{\bf Acknowledgement}\ This work was supported in part by the
Marsden fund of the Royal Society of New Zealand. I acknowledge
helpful discussions and e-mail correspondences with Shin'ichi
Nojiri, Sergei Odintsov and M. Sami during the course of this
project.

\end{document}